# High Temperature Superconductivity in the 2D t-J Model at Metal-Insulator Transition: Variational Mean Field Results

Ken Cappon[*]
*Principia Microsystems Inc., 12 Caines Avenue, Toronto, ON, Canada M2M 1G2*
(2 June 1998)

Conventional Hartree-Fock mean field theory is used, for the first time, to investigate the full two-dimensional t-J model. To date, all other nontrivial mean field approaches modify the Hamiltonian or violate the double occupancy constraint. Unlike conventional mean field theory, these methods do not give variational results for the t-J model.

The mean field phase diagram of the t-J model predicts enhanced superconductivity at the crossover between an insulating phase at low doping and a metallic phase at moderate doping. This behavior is remarkably similar to that of high temperature superconductors, for which the model is expected to provide a good description. In the mean field approximation, the t-J model is described by self-interacting spinless fermions hopping on a background lattice with spiral antiferromagnetic order. The interaction is attractive and high temperature superconductivity arises from the narrow bandwidth at the insulator to metal crossover.

Doping high temperature superconductors introduces charge carriers into copper oxide planes. At low doping these materials are insulators with antiferromagnetic order, while at high doping they are metals. High temperature superconductivity occurs between these limits in the vicinity of the insulator to metal crossover. A temperature-dopant phase diagram for $La_{2-x}Sr_xCuO_4$ is given in Fig. 1. The insulator to metal crossover appears to be related to the phenomenon of high temperature superconductivity. This has been investigated both above and, with a strong magnetic field, below the superconducting transition by Boebringer et al[1].

It is expected that the essential physics of high temperature superconductivity is contained in the t-J model on a square lattice[2]. The t-J model Hamiltonian is defined as

$$H = \sum_{\mathbf{r},\mathbf{a}} \left( \tfrac{1}{2} J \left( \mathbf{S}_{\mathbf{r}} \cdot \mathbf{S}_{\mathbf{r+a}} - \tfrac{1}{4} n_{\mathbf{r}} n_{\mathbf{r+a}} \right) - t \sum_{s} c^{\dagger}_{\mathbf{r+a},s} (1 - n_{\mathbf{r+a}})(1 - n_{\mathbf{r}}) c_{\mathbf{r},s} \right), \qquad (1)$$

where $c^{\dagger}_{\mathbf{r},s}$ creates an electron on site $\mathbf{r} = (n\hat{\mathbf{x}} + m\hat{\mathbf{y}})a$ with spin $s \in \{\uparrow, \downarrow\}$,

$$n_{\mathbf{r}} = \sum_s c^\dagger_{\mathbf{r},s} c_{\mathbf{r},s} \qquad (2)$$

is the number of particles on site r,

$$\mathbf{S}_{\mathbf{r}} \cdot \mathbf{S}_{\mathbf{r+a}} = \tfrac{1}{2}\sum_{s,s'} c^\dagger_{\mathbf{r},s} c^\dagger_{\mathbf{r+a},s'} c_{\mathbf{r+a},s} c_{\mathbf{r},s'} - \tfrac{1}{4} n_{\mathbf{r}} n_{\mathbf{r+a}} \qquad (3)$$

is a nearest neighbor antiferromagnetic interaction and $\mathbf{a} \in \{a\hat{\mathbf{x}}, -a\hat{\mathbf{x}}, a\hat{\mathbf{y}}, -a\hat{\mathbf{y}}\}$ connects nearest neighbors. The Hamiltonian is constrained to the subspace with no double occupancy. For $La_{2-x}Sr_xCuO_4$, the parameter J is approximately 1500°K and the ratio t/J is generally estimated to be in the range of one to five.

An additional assumption that phase separated states are not viable physical configurations will be made about the model. The reason for this is the Coulomb interaction in the t-J model is limited to an onsite term. Although this should be a good approximation for uniform charge distribution, the long range Coulomb interaction will oppose any phase separation with a non-uniform distribution of charge.

The t-J model will be explored with a straightforward application of Hartree-Fock mean field theory. For this development an equivalent form of Hamiltonian will be used

$$H = \lim_{U \to \infty} \sum_{\mathbf{r},\mathbf{a}} \left( \tfrac{1}{2} J (\mathbf{S}_{\mathbf{r}} \cdot \mathbf{S}_{\mathbf{r+a}} - \tfrac{1}{4} n_{\mathbf{r}} n_{\mathbf{r+a}}) - t \sum_s c^\dagger_{\mathbf{r+a},s} c_{\mathbf{r},s} + \tfrac{1}{2} U (n_{\mathbf{r}}^2 - n_{\mathbf{r}}) \right) \qquad (4)$$

where the constraint of no double occupancy is imposed with a repulsive term of infinite energy. Also to balance the conflicting requirements of ferromagnetism favored by the t term and antiferromagnetism favored by the J term, a spiral basis[3] defined by

$$a^\dagger_{\mathbf{r},s} = \cos(\mathbf{q}\cdot\mathbf{r}) c_{\mathbf{r},s} + s\, \sin(\mathbf{q}\cdot\mathbf{r}) c_{\mathbf{r},-s} \qquad (5)$$

will be used. The operator $a_{\mathbf{r},s}$ creates a particle with orientation $s$ at the origin that spirals with spatial frequencies $q_x$ and $q_y$ in the $x$ and $y$ directions respectively. Ferromagnetism corresponds to $q_x = q_y = 0$, and antiferromagnetism to $q_x = q_y = \tfrac{p}{2a}$.

Now for the mean field development the hole density of the up spiral state $\langle a^\dagger_{\mathbf{r},\uparrow} a_{\mathbf{r},\uparrow} \rangle$ is chosen to be less than unity and uniform. Then the onsite repulsion will prevent occupation of the opposite spiral orientation and aside from the repulsive term it will be unnecessary to consider any other terms involving the opposite spiral spin orientation. Identifying the order parameters $\langle a^\dagger_{\mathbf{r+a},\uparrow} a_{\mathbf{r},\uparrow}\rangle$ and $\langle a_{\mathbf{r+a},\uparrow} a_{\mathbf{r},\uparrow}\rangle$ gives the mean field Hamiltonian

$$H_0 = \lim_{U_0 \to \infty} \frac{1}{z}\sum_{\mathbf{r},\mathbf{a}} \begin{pmatrix} m_0\left(a^\dagger_{\mathbf{r},\uparrow} a_{\mathbf{r},\uparrow} - \tfrac{1}{2}\right) + t_0 a^\dagger_{\mathbf{r+a},\uparrow} a_{\mathbf{r},\uparrow} - \tfrac{1}{2}\left(\Delta^*_{\mathbf{a}} a^\dagger_{\mathbf{r+a},\uparrow} a^\dagger_{\mathbf{r},\uparrow} + \Delta_{\mathbf{a}} a_{\mathbf{r+a},\uparrow} a_{\mathbf{r},\uparrow}\right) \\ + U_0 a_{\mathbf{r},\downarrow} a^\dagger_{\mathbf{r},\downarrow} \end{pmatrix} \tag{6}$$

with

$$\Delta_{\mathbf{a}} = -\Delta_{-\mathbf{a}}. \tag{7}$$

The onsite repulsion has limited the degrees of freedom of each site to being empty or occupied by a particle created by $\cos(\mathbf{q}\cdot\mathbf{r}) c^\dagger_{\mathbf{r},\uparrow} + \sin(\mathbf{q}\cdot\mathbf{r}) c^\dagger_{\mathbf{r},\downarrow}$. Thus there is only a single fermion degree of freedom and the system is described by holes hopping in a static background of spiral magnetic order.

Although the impact of magnetic order is to reduce the occupancy of the unfavored spin orientation, in this case the impact is rather severe and it is expected that magnetic order will be emphasized. However, it is very significant that the double occupancy constraint has not been violated. Other mean field approaches to the t-J model modify the Hamiltonian or treat the constraint in some approximate fashion. The penalty for this is energy expectation values for the t-J Hamiltonian cannot be determined. The results are not variational and there is no strict criterion for choosing one configuration over another. Nature at equilibrium rigidly chooses the configuration of lowest free energy.

Now solving the mean field Hamiltonian for the particle number and "order parameters" gives

$$n = \langle a^\dagger_{\mathbf{r}} a_{\mathbf{r}}\rangle = \tfrac{1}{2} - \tfrac{1}{2N}\sum_{\mathbf{k}} \tanh\left(\tfrac{1}{2} b \sqrt{(m_0 + t_{\mathbf{k}})^2 + |\Delta_{\mathbf{k}}|^2}\right) \frac{m_0 + t_{\mathbf{k}}}{\sqrt{(m_0 + t_{\mathbf{k}})^2 + |\Delta_{\mathbf{k}}|^2}} \tag{8}$$

$$p = \langle a^\dagger_{\mathbf{r+a},\uparrow} a_{\mathbf{r},\uparrow}\rangle = -\tfrac{1}{2N}\frac{1}{t_0}\sum_{\mathbf{k}} \tanh\left(\tfrac{1}{2} b \sqrt{(m_0 + t_{\mathbf{k}})^2 + |\Delta_{\mathbf{k}}|^2}\right) \frac{(m_0 + t_{\mathbf{k}}) t_{\mathbf{k}}}{\sqrt{(m_0 + t_{\mathbf{k}})^2 + |\Delta_{\mathbf{k}}|^2}} \tag{9}$$

$$s_{\mathbf{d}} = \langle a_{\mathbf{r+a},\uparrow} a_{\mathbf{r},\uparrow}\rangle = -\tfrac{1}{2N}\frac{1}{\Delta_{\mathbf{a}}}\sum_{\mathbf{k}} \tanh\left(\tfrac{1}{2} b \sqrt{(m_0 + t_{\mathbf{k}})^2 + |\Delta_{\mathbf{k}}|^2}\right) \frac{|\Delta_{\mathbf{k}}|^2}{\sqrt{(m_0 + t_{\mathbf{k}})^2 + |\Delta_{\mathbf{k}}|^2}} \tag{10}$$

$$\langle a^\dagger_{\mathbf{r+a},\downarrow} a_{\mathbf{r},\downarrow}\rangle = \langle a_{\mathbf{r+a},\downarrow} a_{\mathbf{r},\downarrow}\rangle = \langle a^\dagger_{\mathbf{r+a},\downarrow} a_{\mathbf{x},\uparrow}\rangle = \langle a_{\mathbf{r+a},\downarrow} a_{\mathbf{r},\uparrow}\rangle = \langle a_{\mathbf{r},\downarrow} a_{\mathbf{r},\uparrow}\rangle = 0 \tag{11}$$

where

$$t_\mathbf{k} = \tfrac{1}{z} t_0 \sum_\mathbf{a} \cos(\mathbf{k}\cdot\mathbf{a}), \qquad \Delta_\mathbf{k} = \tfrac{1}{z}\sum_\mathbf{a} \Delta_\mathbf{a} \sin(\mathbf{k}\cdot\mathbf{a}) \qquad (12)$$

and the mean field expectation value for the operator $A$ is defined as $\langle A \rangle = \dfrac{\mathrm{Tr}\langle A e^{-bH_0}\rangle}{\mathrm{Tr}\langle e^{-bH_0}\rangle}$.

In addition the mean field free energy is

$$F_0 = -\tfrac{1}{b}\sum_\mathbf{k} \ln\left(2\cosh\left(\tfrac{1}{2} b \sqrt{(m_0 + t_\mathbf{k})^2 + |\Delta_\mathbf{k}|^2}\right)\right). \qquad (13)$$

Mean field theory is based on the fact that the free energy $F = -\tfrac{1}{b}\ln(\mathrm{Tr}\langle e^{-bH}\rangle)$ is bounded above by $F_{MF} = \langle H \rangle - \langle H_0 \rangle + F_0$. This is true for <u>any</u> Hamiltonian $H_0$ provided of course that it operates in the same space as $H$. However, the mean field Hamiltonian $H_0$ is chosen to have only single particle interactions so that it is solvable. Mean field theory can also be presented using, and is equivalent to, a Hubbard-Stratonovitch decoupling and steepest descent or saddle point approximation.

One is free to choose the gap symmetry and spiral spin arrangement. The gap symmetry $\Delta_\mathbf{a} = \Delta(\mathrm{sgn}(a_x) + i\,\mathrm{sgn}(a_y))$ and spiral spin arrangement $g_x = g_y = g$ have been found optimal. With this choice,

$$\Delta_\mathbf{k} = \tfrac{1}{2}\Delta(\sin(k_x a) + i\sin(k_y a)). \qquad (14)$$

Also in the same notation

$$t_\mathbf{k} = \tfrac{1}{2} t_0 (\cos(k_x a) + \cos(k_y a)), \qquad (15)$$

and

$$\tfrac{1}{N}\langle H \rangle = -\tfrac{z}{4} J \sin^2(qa)\left((1-n)^2 - p^2 + s^2\right) - zt\cos(qa)p, \qquad (16)$$

$$\tfrac{1}{N}\langle H_0 \rangle = t_0 p + m_0(n - \tfrac{1}{2}) - \Delta s. \qquad (17)$$

Fixing the number of particles $n$ and minimizing the free energy bound $F_{MF}$ with respect to the applied fields $t_0$ and $\Delta$, gives the "self-consistent" relations

$$t_0 = \tfrac{z}{2}(2t\cos(qa) - J\sin^2(qa)p), \qquad (18)$$

$$\Delta = \tfrac{z}{2} J\sin^2(qa) s. \qquad (19)$$

Minimizing with respect to $q$ gives

$$\frac{J}{2t}\cos(qa) = \frac{p}{(1-n)^2 - p^2 + s^2}. \tag{20}$$

There are two possible "broken symmetries," the gap and the bandwidth. The bandwidth is an unusual broken symmetry that arises because antiferromagnetic ordering impedes conduction. Thus, there are four competing phases:

- Insulating phase, where the gap $\Delta$ is non zero, and the bandwidth $t_0$ is zero. It has strict antiferromagnetic order, charge carrier pairing and no conductivity.
- Metallic phase, where the bandwidth is nonzero and the gap is zero. There is a Fermi surface and conduction.
- Superconducting phase, where both the gap and bandwidth are nonzero.
- High temperature insulating phase, where both the gap and bandwidth are zero.

Finding the phase of lowest free energy gives the temperature-dopant phase diagrams of Fig. 3 for $t/J = 1.1, 1.0, 0.9$. At low doping the system is an insulator and at high doping it is metallic. At intermediate doping there is a region of enhanced superconductivity in the vicinity of the insulator to metal crossover. The values of t/J were chosen to give an insulator to metal transition at near to the same doping as $La_{2-x}Sr_xCuO_4$. Larger t/J gives the crossover at a lower doping and a lower transition temperature, smaller t/J gives a higher transition temperature and higher doping at the crossover.

For completeness, the phase diagram for t/J = 1 is given in Fig. 4 for a wider range of temperatures[4].

To help understand the interplay between phases, it is instructive to consider the simpler case of zero temperature and low doping. At optimum q, the energy is then given by

$$\tfrac{1}{N}\langle H \rangle = -\tfrac{z}{4}J\left((1-n)^2 + s^2 + \left(4\frac{t^2}{J^2} - 1\right)p^2 + O(n^3)\right). \tag{21}$$

For a given hole density $n$ there is a tradeoff between the size of the spin gap and the bandwidth. At low doping the hole density has the dependence $n = O(p) + O(s^2)$ and the energy is minimized with a spin gap ($s > 0$) and zero bandwidth ($p = 0$). With increased doping there is a first order phase transition from the insulating to superconducting phase when the bandwidth takes on a nonzero value. Further doping decreases the gap and the bandwidth increases. The superconducting phase has the largest gap at the transition from the insulating phase where the bandwidth is narrowest.

The mean field phase diagram is similar to that of $La_{2-x}Sr_xCuO_4$. Furthermore, the insulating phase is predicted to have a gap, which is observed in many high temperature superconductors[5]. However, mean field theory gives an idealized picture of the system and cannot be expected to precisely reproduce the experimental phase diagram. Rather one hopes to get a qualitative picture and capture the essential physics. One must also bear in mind that the t-J model itself is an approximate description of high temperature superconductors, indeed not all superconductors believed to be modeled by the t-J model have identical phase diagrams.

In the mean field description, superconductivity peaks at metal insulator crossover whereas in $La_{2-x}Sr_xCuO_4$ it peaks at higher doping. This might be explained by reduction of the transition temperature due to phase fluctuations of the gap order parameter[6] at low doping which cannot be incorporated into the mean field description. Another discrepancy between theory and experiment is that the insulating phase has perfect antiferromagnetic order and no conduction. Experimentally the insulating phase has some conduction with the resistivity increasing with decreasing temperature. It is also known the t-J model itself will support some conduction[7]. As far as magnetic order, $La_{2-x}Sr_xCuO_4$ has Néel order only at very low doping, although antiferromagnetic order persists beyond this it is no longer macroscopic. The mean field approximation not only has perfect order but infinite range. This and the lack of conduction are likely due to an overemphasis of magnetic order in the mean field approximation and the absence of fluctuations of the spin background.

There has been debate about whether high temperature superconductors have a gap of s-wave or d-wave symmetry. The mean field gap contains both. The spatially averaged gap has the symmetry $(1+i)s + (1-i)d_{x^2-y^2}$. The gap order parameter is given by

$$\tfrac{1}{2}\langle c_{\mathbf{k},\uparrow}c_{-\mathbf{k},\downarrow} - c_{\mathbf{k},\downarrow}c_{-\mathbf{k},\uparrow}\rangle = -\tfrac{1}{4}\left(\langle a^\dagger_{\mathbf{k+q},\uparrow}a^\dagger_{-\mathbf{k-q},\uparrow}\rangle + \langle a^\dagger_{-\mathbf{k+q},\uparrow}a^\dagger_{\mathbf{k-q},\uparrow}\rangle\right) \tag{22}$$

where

$$\langle a^\dagger_{\mathbf{k},\uparrow}a^\dagger_{-\mathbf{k},\uparrow}\rangle = -\tfrac{1}{2N}\tanh\left(\tfrac{1}{2}\mathbf{b}\sqrt{(\mathbf{m}_0 + t_\mathbf{k})^2 + |\Delta_\mathbf{k}|^2}\right)\frac{\Delta_\mathbf{k}}{\sqrt{(\mathbf{m}_0 + t_\mathbf{k})^2 + |\Delta_\mathbf{k}|^2}}. \tag{23}$$

This follows from the relation $c_{\mathbf{k},\uparrow} = \frac{1}{2}\left(a^\dagger_{\mathbf{k+q},\uparrow} + s\, a^\dagger_{\mathbf{k-q},\uparrow}\right)$.

The other non-zero independent gap parameters are

$$\left\langle c_{\mathbf{k},s}\, c_{-\mathbf{k},s}\right\rangle = \tfrac{1}{4} s \left( \left\langle a^\dagger_{\mathbf{k+q},\uparrow} a^\dagger_{-\mathbf{k-q},\uparrow}\right\rangle - \left\langle a^\dagger_{-\mathbf{k+q},\uparrow} a^\dagger_{\mathbf{k-q},\uparrow}\right\rangle \right) \tag{24}$$

and

$$\left\langle c_{\mathbf{k\pm q},s}\, c_{-\mathbf{k\pm q},s}\right\rangle = \mp \tfrac{1}{2}\left\langle c_{\mathbf{k\pm q},\uparrow} c_{-\mathbf{k\pm q},\downarrow} + c_{\mathbf{k\pm q},\downarrow} c_{-\mathbf{k\pm q},\uparrow}\right\rangle = \tfrac{1}{4}\left\langle a^\dagger_{\mathbf{k},\uparrow} a^\dagger_{-\mathbf{k},\uparrow}\right\rangle . \tag{25}$$

The mean field approximation depends only on the relative orientation of the spin of nearest neighbors and therefore the spiral spin arrangement is degenerate with the double spiral and canting arrangements[8]. The mean field development is no different but the gap functions differ due to the dependence of $c_{\mathbf{k},s}$ on $a_{\mathbf{k},\uparrow}$. The gap functions are equally complex and contain both s-wave and d-wave symmetry.

Mean field theory has played an important role in physics. It gives a core understanding of many physical systems and collective phenomena. The reason for this is not fully understood, but mean field theory allows one to work with infinite size systems, which is essential to investigate phase transitions. Also, mean field theory uses the same criteria, minimum free energy, used by nature to choose the system state. Although mean field theory does not have access to the full spectrum of wavefunctions as nature does, the spectrum is often rich enough to understand much of the physics involved.

Historically it has usually taken a mean field description of a Hamiltonian before there is consensus for its phase diagram, albeit with caveats and "corrections." The mean field theory of the t-J model developed here shows interplay between insulating and metallic phases and how superconductivity with an enhanced transition temperature arises at the insulator to metal crossover.

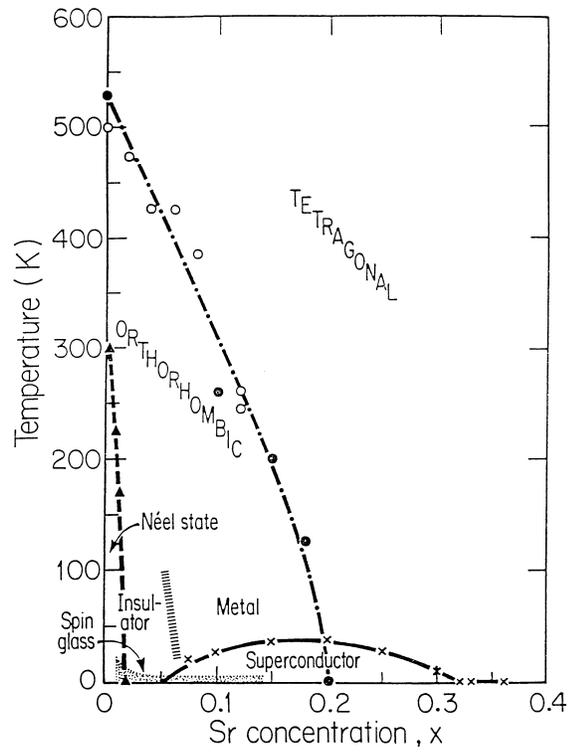

FIG. 1 Phase diagram of $La_{2-x}Sr_xCuO_4$. Note the metal insulator crossover (wide hatched line) and its proximity to the superconducting phase. From Birgeneau and Shirane[9].

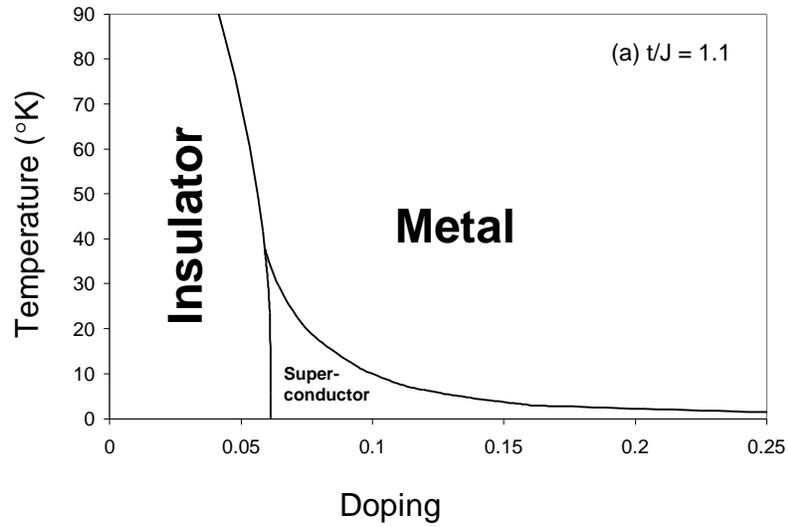
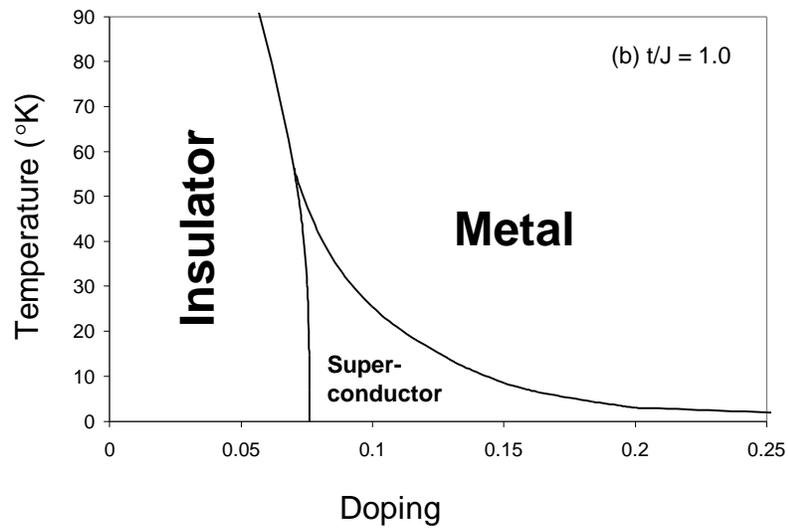
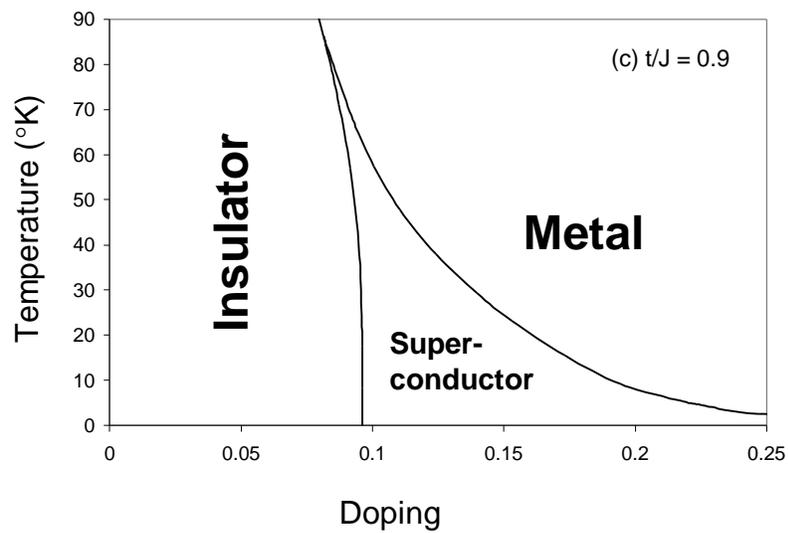

FIG. 3. Low temperature mean field phase diagrams for $J = 1500°K$, $t/J = 1.1, 1.0, 0.9$.

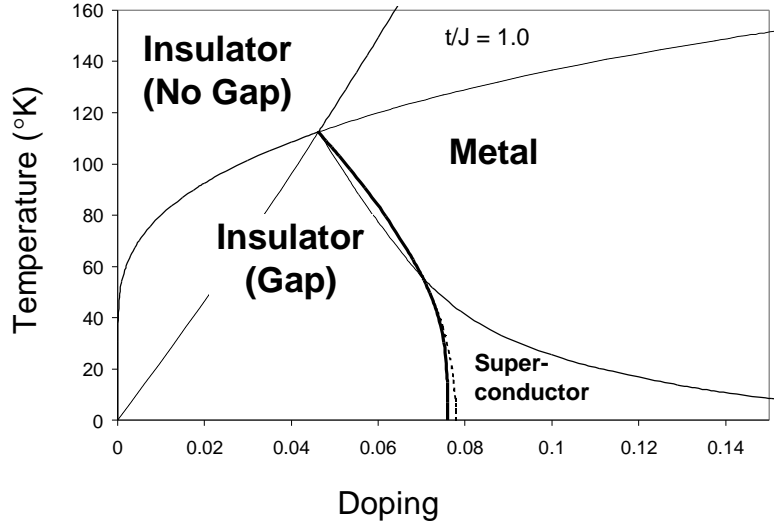

FIG. 4. Mean field phase diagram for J = 1500°K, t/J = 1.0. The solid lines denote the phase boundaries. To provide insight into mechanisms involved, the dashed lines extend the boundaries into regions in which the bounded phases are no longer stable. These boundaries only exist if the phase in which they are contained is suppressed. For example, the metal-insulator(gap) boundary within the superconducting phase represents the metal-insulator transition if superconductivity is suppressed.

The bold lines denote first order phase transitions, the remainder are second order transitions.

---

[*] email: kenc@princip.com

[1] G. S. Boebinger, Y. Ando, A. Passner, T. Kimura, M. Okuya, J. Shimoyama, K. Kishio, K. Tamasuku, N. Ichikawa, and S. Uchida, Phys. Rev. Lett. **77**, 5417 (1996)

[2] F. C. Zhang and T. M. Rice, Phys. Rev. B **37**, 3759 (1988)

[3] B. Shraiman and E. Siggia, Phys. Rev. Lett **62**, 1564 (1989); C. L. Kane, P. A. Lee, T. K. Ng, B. Chakraborty and N. Read, Phys. Rev. B **41**, 2653 (1990)

[4] The high temperature insulating phase is in fact unstable to the usual mean field description of the Heisenberg antiferromagnet with static holes. This phase is also insulating with no gap or bandwidth and results in only slight movement of the insulating (no gap) phase boundary.

[5] A. G. Loeser, Z.-X. Shen, D. S. Dessau, D. S. Marshall. C. H. Park, P. Fournier and A. Kapitulnik, Science **273**, 325 (1996)

[6] V. J. Emery and S. A. Kivelson, Nature **374**, 434 (1995)

[7] S. A. Trugman, Phys. Rev. B **37**, 1597, (1988)

[8] defined in Ref. 3b.

[9] R. J. Birgeneau and G. Shirane, in *Physical Properties of High Temperature Superconductors I,* edited by D. M. Ginsberg (World Scientific, Singapore, 1989).